\documentstyle[12pt]{article}
\newlength{\extraspace}
\newlength{\extraspaces}
\newcommand{\be}{\begin{equation}
\addtolength{\abovedisplayskip}{\extraspaces}
\addtolength{\belowdisplayskip}{\extraspaces}
\addtolength{\abovedisplayshortskip}{\extraspace}
\addtolength{\belowdisplayshortskip}{\extraspace}}
\newcommand{\ee}{\end{equation}}
\newcommand{\ba}{\begin{eqnarray}
\addtolength{\abovedisplayskip}{\extraspaces}
\addtolength{\belowdisplayskip}{\extraspaces}
\addtolength{\abovedisplayshortskip}{\extraspace}
\addtolength{\belowdisplayshortskip}{\extraspace}}
\newcommand{\ea}{\end{eqnarray}}
\newcommand{\nonu}{\nonumber \\[.5mm]}
\newcommand{\A}{&\!\!\!}
\begin{document}
\thispagestyle{empty}
\begin{flushright}
SIT-LP-08/10 \\
October, 2008
\end{flushright}
\vspace{7mm}
%
%
\begin{center}
{\large{\bf $N = 2$ SUSY QED in nonlinear/linear SUSY relation}} \\[20mm]
{\sc Kazunari Shima}
\footnote{
\tt e-mail: shima@sit.ac.jp} \ 
and \ 
{\sc Motomu Tsuda}
\footnote{
\tt e-mail: tsuda@sit.ac.jp} 
\\[5mm]
{\it Laboratory of Physics, 
Saitama Institute of Technology \\
Fukaya, Saitama 369-0293, Japan} \\[20mm]
\begin{abstract}
We show systematically the relation between a $N = 2$ nonlinear supersymmetric (NLSUSY) model 
and a $N = 2$ SUSY QED theory by means of the superfield formulation in two dimensional spacetime 
without imposing a priori any special gauge conditions for a gauge superfield. 
\\[5mm]
\noindent
PACS: 11.30.Pb, 12.60.Jv, 12.60.Rc, 12.10.-g \\[2mm]
\noindent
Keywords: supersymmetry, superfield, Nambu-Goldstone fermion, 
nonlinear/linear SUSY relation, composite unified theory 
\end{abstract}
\end{center}

\newpage

\noindent
{\it Nonlinear/linear supersymmetry (NL/L SUSY) relation}, i.e. various LSUSY theories 
(with spontaneous SUSY breaking) represented in a NLSUSY-model background is important to investigate 
the low energy particle physics of NLSUSY general relativity (GR) \cite{KS} in the SGM scenario \cite{ST1}. 
In the NL/L SUSY relation L supermultiplets are realized as the (massless) eigenstates 
of spacetime symmetry in terms of SUSY composites of Nambu-Goldstone (NG) fermions, 
which are called {\it SUSY invariant relations}. 
Recently, the relation between the NLSUSY model \cite{VA} 
and {\it interacting} LSUSY (Yukawa-interaction and QED) theories has been shown \cite{ST2,ST3} 
for (realistic) $N = 2$ SUSY in two dimensional spacetime ($d = 2$) 
through the SUSY invariant relations, 
which gives new insights into the cosmology and the low energy particle physics of NLSUSY GR \cite{STL}. 
Furthermore, the $N = 2$ LSUSY Yukawa-interaction theory (for a vector supermultiplet) 
in NL/L SUSY relation has been discussed systematically by means of the $d = 2$ superfield formulation 
without imposing any special gauge conditions for a general gauge superfield \cite{ST4}. 

In this letter we show {\it systematically} the relation between the $N = 2$ NLSUSY model 
and the $N = 2$ SUSY QED theory by using a $d = 2$, $N = 2$ general gauge superfield \cite{DVF,ST5} 
and without imposing a priori any specific SUSY gauge conditions in contrast with Ref.\cite{IK1}. 
Also, it is crucial for the SGM scenario to attribute straightforwardly LSUSY actions 
to a NLSUSY action \cite{VA} alone and $N = 2$ SUSY is the minimal and realistic one in NLSUSY GR. 

Let us introduce $N = 2$ SUSY invariant relations for vector and scalar supermultiplets in $d = 2$. 
Those are obtained from the superfield formulation as briefly explains below. 
The $N = 2$ general gauge \cite{DVF,ST5} and the $N = 2$ scalar (for example, see \cite{UZ}) superfields 
on superspace coordinates $(x^a, \theta_\alpha^i)$ ($i = 1, 2$) are given respectively by 
%
%
%
%
\ba
{\cal V}(x, \theta) \A = \A C(x) + \bar\theta^i \Lambda^i(x) 
+ {1 \over 2} \bar\theta^i \theta^j M^{ij}(x) 
- {1 \over 2} \bar\theta^i \theta^i M^{jj}(x) 
+ {1 \over 4} \epsilon^{ij} \bar\theta^i \gamma_5 \theta^j \phi(x) 
\nonu
\A \A 
- {i \over 4} \epsilon^{ij} \bar\theta^i \gamma_a \theta^j v^a(x) 
- {1 \over 2} \bar\theta^i \theta^i \bar\theta^j \lambda^j(x) 
- {1 \over 8} \bar\theta^i \theta^i \bar\theta^j \theta^j D(x), 
\label{VSF}
\\
\Phi^i(x, \theta) \A = \A B^i(x) + \bar\theta^i \chi(x) - \epsilon^{ij} \bar\theta^j \nu(x) 
- {1 \over 2} \bar\theta^j \theta^j F^i(x) + \bar\theta^i \theta^j F^j(x) 
- i \bar\theta^i \!\!\not\!\partial B^j(x) \theta^j 
\nonu
\A \A 
+ {i \over 2} \bar\theta^j \theta^j (\bar\theta^i \!\!\not\!\partial \chi(x) 
- \epsilon^{ik} \bar\theta^k \!\!\not\!\partial \nu(x)) 
+ {1 \over 8} \bar\theta^j \theta^j \bar\theta^k \theta^k \Box B^i(x), 
\label{SSF}
\ea
where we denote the component fields in the gauge superfield (\ref{VSF}) 
by $(C, D)$ for two scalar fields, $(\Lambda^i, \lambda^i)$ for four (Majorana) spinor fields, 
$\phi$ for a pseudo scalar field, $v^a$ for a vector field, 
and $M^{ij} = M^{(ij)}$ $\left(= {1 \over 2}(M^{ij} + M^{ji}) \right)$ 
for three scalar fields ($M^{ii} = \delta^{ij} M^{ij}$), 
while in the scalar superfields (\ref{SSF}) by $B^i$ for two scalar fields, 
$(\chi, \nu)$ for two (Majorana) spinor fields and $F^i$ for two auxiliary scalar fields. 
Defining the following specific supertranslations \cite{UZ,IK2} 
of superspace coordinates $(x^a, \theta^i)$ depending on the (Majorana) NG fermions $\psi^i$ 
in the NLSUSY model, 
\ba
\A \A 
x'^a = x^a + i \kappa \bar\theta^i \gamma^a \psi^i, 
\nonu
\A \A 
\theta'^i = \theta^i - \kappa \psi^i, 
\label{specific}
\ea
with the constant $\kappa$ whose dimension is (mass)$^{-1}$, 
and considering the $N = 2$ superfields (\ref{VSF}) and (\ref{SSF}) on the specific coordinates (\ref{specific}), 
i.e. 
\be
\tilde{\cal V}(x, \theta) = {\cal V}(x', \theta'), 
\ \ \ \tilde \Phi^i(x, \theta) = \Phi^i(x', \theta'), 
\label{SFpsi}
\ee
lead to the SUSY invariant relations. 
Indeed, the superfields (\ref{SFpsi}) can be expanded in the power series of $\theta^i$ as 
\ba
\tilde{\cal V}(x, \theta) 
\A = \A \tilde C(x) + \bar\theta^i \tilde \Lambda^i(x) 
+ {1 \over 2} \bar\theta^i \theta^j \tilde M^{ij}(x) 
- {1 \over 2} \bar\theta^i \theta^i \tilde M^{jj}(x) 
+ {1 \over 4} \epsilon^{ij} \bar\theta^i \gamma_5 \theta^j \tilde \phi(x) 
\nonu
\A \A 
- {i \over 4} \epsilon^{ij} \bar\theta^i \gamma_a \theta^j \tilde v^a(x) 
- {1 \over 2} \bar\theta^i \theta^i \bar\theta^j \tilde \lambda^j(x) 
- {1 \over 8} \bar\theta^i \theta^i \bar\theta^j \theta^j \tilde D(x), 
\label{VSFtilde}
\\
\tilde \Phi^i(x, \theta) 
\A = \A \tilde B^i(x) + \bar\theta^i \tilde \chi(x) - \epsilon^{ij} \bar\theta^j \tilde \nu(x) 
- {1 \over 2} \bar\theta^j \theta^j \tilde F^i(x) + \bar\theta^i \theta^j \tilde F^j(x) + \cdots, 
\label{SSFtilde}
\end{eqnarray}
where the component fields $(\tilde C(x), \tilde\Lambda^i(x), \tilde M^{ij}(x), \cdots)$ 
and $(\tilde B^i(x), \tilde\chi(x), \tilde\nu(x), \tilde F^i(x)$) 
are expressed in terms of $\psi^i$ and the initial component fields in Eqs.(\ref{VSF}) and (\ref{SSF}), 
whose explicit forms are given in Ref.\cite{ST6} for Eq.(\ref{VSFtilde}), 
while those for Eq.(\ref{SSFtilde}) are obtained by straightforward calculations as \cite{UZ} 
\ba
\tilde B^i \A = \A B^i - \kappa (\bar\psi^i \chi - \epsilon^{ij} \bar\psi^j \nu) 
- {1 \over 2} \kappa^2 (\bar\psi^j \psi^j F^i - 2 \bar\psi^i \psi^j F^j 
+ 2 i \bar\psi^i \!\!\not\!\partial B^j \psi^j) 
\nonu
\A \A 
- {i \over 2} \kappa^3 \bar\psi^j \psi^j (\bar\psi^i \!\!\not\!\partial \chi 
- \epsilon^{ik} \bar\psi^k \!\!\not\!\partial \nu) 
+ {1 \over 8} \kappa^4 \bar\psi^j \psi^j \bar\psi^k \psi^k \Box B^i, 
\nonu
\tilde\chi \A = \A \chi - \kappa (\psi^i F^i - i \!\!\not\!\partial B^i \psi^i) 
+ {i \over 2} \kappa^2 \{ \not\!\partial \chi \bar\psi^i \psi^i 
- \epsilon^{ij} (\psi^i \bar\psi^j \!\!\not\!\partial \nu - \gamma^a \psi^i \bar\psi^j \partial_a \nu) \} 
\nonu
\A \A 
- {1 \over 2} \kappa^3 \psi^i \bar\psi^j \psi^j \Box B^i 
- {i \over 2} \kappa^3 \!\!\not\!\partial F^i \psi^i \bar\psi^j \psi^j 
- {1 \over 8} \kappa^4 \Box \chi \bar\psi^i \psi^i \bar\psi^j \psi^j, 
\nonu
\tilde\nu \A = \A \nu + \kappa \epsilon^{ij} (\psi^i F^j - i \!\!\not\!\partial B^i \psi^j) 
+ {i \over 2} \kappa^2 \{ \not\!\partial \nu \bar\psi^i \psi^i 
+ \epsilon^{ij} (\psi^i \bar\psi^j \!\!\not\!\partial \chi - \gamma^a \psi^i \bar\psi^j \partial_a \chi) \} 
\nonu
\A \A 
- {1 \over 2} \kappa^3 \epsilon^{ij} \psi^i \bar\psi^k \psi^k \Box B^j 
- {i \over 2} \kappa^3 \epsilon^{ij} \!\!\not\!\partial F^i \psi^j \bar\psi^k \psi^k 
- {1 \over 8} \kappa^4 \Box \nu \bar\psi^i \psi^i \bar\psi^j \psi^j, 
\nonu
\tilde F^i \A = \A F^i + i \kappa (\bar\psi^i \!\!\not\!\partial \chi 
+ \epsilon^{ij} \bar\psi^j \!\!\not\!\partial \nu) 
+ {1 \over 2} \kappa^2 \bar\psi^j \psi^j \Box B^i - \kappa^2 \bar\psi^i \psi^j \Box B^j 
- i \kappa^2 \bar\psi^i \!\!\not\!\partial F^j \psi^j 
\nonu
\A \A 
- {1 \over 2} \kappa^3 \bar\psi^j \psi^j (\bar\psi^i \Box \chi + \epsilon^{ik} \bar\psi^k \Box \nu) 
+ {1 \over 8} \kappa^4 \bar\psi^j \psi^j \bar\psi^k \psi^k \Box F^i. 
\ea
Then, we can impose the (simplest) constraints, 
which are {\it SUSY invariant} \cite{UZ,IK2} and eliminate other degrees of freedom than $\psi^i$, 
as follows; 
\ba
\A \A 
\tilde C = \tilde\Lambda^i = \tilde M^{ij} = \tilde\phi = \tilde v^a = \tilde\lambda^i = 0, 
\ \ \ \tilde D = {\xi \over \kappa}, 
\label{constraintI}
\\
\A \A 
\tilde B^i = \tilde\chi = \tilde\nu = 0, 
\ \ \ \tilde F^i = {\xi^i \over \kappa}, 
\label{constraintII}
\ea
with arbitrary real paramaters $\xi$ and $\xi^i$. 
By solving the SUSY invariant constraints (\ref{constraintI}), 
the SUSY invariant relations for the $N = 2$ vector supermultiplet are given in all orders of $\psi^i$ as 
\cite{ST6} 
\ba
C \A = \A - {1 \over 8} \xi \kappa^3 \bar\psi^i \psi^i \bar\psi^j \psi^j \vert w \vert, 
\nonu
\Lambda^i \A = \A - {1 \over 2} \xi \kappa^2 
\psi^i \bar\psi^j \psi^j \vert w \vert, 
\nonu
M^{ij} \A = \A {1 \over 2} \xi \kappa \bar\psi^i \psi^j \vert w \vert, 
\nonu
\phi \A = \A - {1 \over 2} \xi \kappa \epsilon^{ij} \bar\psi^i \gamma_5 \psi^j \vert w \vert, 
%
\nonu
v^a \A = \A - {i \over 2} \xi \kappa \epsilon^{ij} \bar\psi^i \gamma^a \psi^j \vert w \vert, 
\nonu
\lambda^i \A = \A \xi \psi^i \vert w \vert, 
\nonu
D \A = \A {\xi \over \kappa} \vert w \vert, 
\label{VSUSYinv}
\end{eqnarray}
while by solving Eq.(\ref{constraintII}) the SUSY invariant relations for the $N = 2$ scalar supermultiplet 
can be calculated in all orders of $\psi^i$, whose explicit forms have been obtained (heuristicly) 
in Ref.\cite{ST2} as 
\ba
\chi \A = \A \xi^i \left[ \psi^i \vert w \vert
+ {i \over 2} \kappa^2 \partial_a 
( \gamma^a \psi^i \bar\psi^j \psi^j \vert w \vert 
) \right], 
\nonu
B^i \A = \A - \kappa \left( {1 \over 2} \xi^i \bar\psi^j \psi^j 
- \xi^j \bar\psi^i \psi^j \right) \vert w \vert, 
\nonu
\nu \A = \A \xi^i \epsilon^{ij} \left[ \psi^j \vert w \vert 
+ {i \over 2} \kappa^2 \partial_a 
( \gamma^a \psi^j \bar\psi^k \psi^k \vert w \vert 
) \right], 
\nonu
F^i \A = \A {1 \over \kappa} \xi^i \left\{ \vert w \vert 
+ {1 \over 8} \kappa^3 
\partial_a \partial^a ( \bar\psi^j \psi^j \bar\psi^k \psi^k \vert w \vert ) 
\right\} 
\nonu
\A \A 
- i \kappa \xi^j \partial_a ( \bar\psi^i \gamma^a \psi^j \vert w \vert ). 
\label{SSUSYinv}
\end{eqnarray}
Here we give Eqs.(\ref{VSUSYinv}) and (\ref{SSUSYinv}) 
as the form containing some vanishing terms due to $(\psi^i)^5 \equiv 0$, 
and $\vert w \vert$ is the determinant \cite{VA} which induces a spacetime-volume differential form 
describing the dynamics of $\psi^i$, i.e. in $d = 2$, 
\ba
\vert w \vert \A = \A \det(w^a{}_b) = \det(\delta^a_b + t^a{}_b) 
\nonu
\A = \A 1 + t^a{}_a + {1 \over 2!}(t^a{}_a t^b{}_b - t^a{}_b t^b{}_a) 
\nonu
\A = \A 1 - i \kappa^2 \bar\psi^i \!\!\not\!\partial \psi^i 
- {1 \over 2} \kappa^4 
(\bar\psi^i \!\!\not\!\partial \psi^i \bar\psi^j \!\!\not\!\partial \psi^j 
- \bar\psi^i \gamma^a \partial_b \psi^i \bar\psi^j \gamma^b \partial_a \psi^j) 
\nonu
\A = \A 1 - i \kappa^2 \bar\psi^i \!\!\not\!\partial \psi^i 
- {1 \over 2} \kappa^4 \epsilon^{ab} 
(\bar\psi^i \psi^j \partial_a \bar\psi^i \gamma_5 \partial_b \psi^j 
+ \bar\psi^i \gamma_5 \psi^j \partial_a \bar\psi^i \partial_b \psi^j) 
\ea
with $t^a{}_b = - i \kappa^2 \bar\psi^i \gamma^a \partial_b \psi^i$. 

Let us discuss below the relation between a $N = 2$ NLSUSY action and a $N = 2$ SUSY QED one 
under the SUSY invariant relations (\ref{VSUSYinv}) and (\ref{SSUSYinv}) in the superfield formulation. 
The $N = 2$ NLSUSY action is given by \cite{VA} 
\be
S_{N = 2{\rm NLSUSY}} = - {1 \over {2 \kappa^2}} \int d^2 x \ \vert w \vert. 
\label{NLSUSYaction}
\ee
On the other hand, the general $N = 2$ SUSY QED action constructed from the superfields 
(\ref{VSF}) and (\ref{SSF}) is defined as 
\be
S^{\rm gen.}_{N = 2{\rm SUSYQED}} 
= S^{\rm gen.}_{{\cal V}{\rm kin}} + S^{\rm gen.}_{{\cal V}{\rm FI}} 
+ S^{\rm gen.}_{\Phi{\rm kin}} + S^{\rm gen.}_e 
\label{SQEDaction}
\ee
with 
\ba
\A \A 
S^{\rm gen.}_{{\cal V}{\rm kin}} 
= {1 \over 32} \int d^2 x \left\{ \int d^2 \theta^i 
\ (\overline{D^i {\cal W}^{jk}} D^i {\cal W}^{jk} 
+ \overline{D^i {\cal W}_5^{jk}} D^i {\cal W}_5^{jk}) \right\}_{\theta^i = 0}, 
\label{Vkin}
\\
\A \A 
S^{\rm gen.}_{{\cal V}{\rm FI}} 
= {\xi \over {2 \kappa}} \int d^2 x \int d^4 \theta^i \ {\cal V}, 
\label{VFI}
\\
\A \A 
S^{\rm gen.}_{\Phi{\rm kin}} + S^{\rm gen.}_e 
= - {1 \over 16} \int d^2 x \int d^4 \theta^i \ e^{-4e{\cal V}} (\Phi^j)^2, 
\label{gauge}
\ea
where we denote the kinetic terms for the vector supermultiplet by $S^{\rm gen.}_{{\cal V}{\rm kin}}$, 
the Fayet-Iliopoulos (FI) term by $S^{\rm gen.}_{{\cal V}{\rm FI}}$, 
and the kinetic terms for the matter scalar supermultiplet 
and the gauge interaction terms by $S^{\rm gen.}_{\Phi{\rm kin}} + S^{\rm gen.}_e$, respectively. 
In Eq.(\ref{Vkin}), the scalar and pseudo scalar superfields, ${\cal W}^{ij}$ and ${\cal W}_5^{ij}$, are 
\be
{\cal W}^{ij} = \bar D^i D^j {\cal V}, \ \ \ {\cal W}_5^{ij} = \bar D^i \gamma_5 D^j {\cal V} 
\ee
with $D^i = {\partial \over \partial\bar\theta^i} - i \!\!\not\!\partial \theta^i$. 
In Eq.(\ref{gauge}), $e$ means a gauge coupling constant whose dimension is $({\rm mass})^1$ in $d = 2$. 

By changing the integration variables in Eq.(\ref{SQEDaction}) from $(x, \theta^i)$ to $(x', \theta'^i)$ 
of Eq.(\ref{specific}) under the SUSY invariant constraints (\ref{constraintI}) and (\ref{constraintII}), 
we can show that 
\ba
\A \A 
(S^{\rm gen.}_{{\cal V}{\rm kin}} +  S^{\rm gen.}_{{\cal V}{\rm FI}})(\psi) = \xi^2 S_{N = 2{\rm NLSUSY}}, 
\nonu
\A \A 
(S^{\rm gen.}_{\Phi{\rm kin}} + S^{\rm gen.}_e)(\psi) = - (\xi^i)^2 S_{N = 2{\rm NLSUSY}}, 
\ea
where we have used a Jacobian calculated as \cite{ST6} 
\be
J(x, \theta^i) = \vert w \vert \det(\delta_b^a - i \kappa \nabla_b \bar\psi^i \gamma^a \theta^i) 
\ee
with $\nabla_a = (w^{-1})_a{}^b \partial_b$. 
Therefore, $N = 2$ SUSY QED action (\ref{SQEDaction}) exactly reduces 
to the $N = 2$ NLSUSY action (\ref{NLSUSYaction}), i.e. 
\be
S^{\rm gen.}_{N = 2{\rm SUSYQED}}(\psi) = S_{N = 2{\rm NLSUSY}}, 
\label{SQED-NLSUSY}
\ee
when $\xi^2 - (\xi^i)^2 = 1$. 
Note that the FI term (\ref{VFI}) indicating spontaneous SUSY breaking 
gives the correct sign of the NLSUSY action in the relation (\ref{SQED-NLSUSY}). 
Also, the gauge interaction terms in Eq.(\ref{gauge}) vanish (nontrivially) in NL/L SUSY relation, 
i.e. 
\be
S^{\rm gen.}_e(\psi)\ {\rm at}\ {\cal O}(e) 
= {1 \over 4} \int d^2 x \int d^4 \theta^i \ e{\cal V} (\Phi^j)^2 (\psi) = 0, 
\label{gauge-vanish}
\ee
and the higher order terms in $S^{\rm gen.}_e(\psi)$ at ${\cal O}(e^n)$ ($n \ge 2$) 
vanish due to $(\psi^i)^5 \equiv 0$. 

Here we explain the relation (\ref{gauge-vanish}) in the explicit component form. 
The gauge interaction terms at ${\cal O}(e)$ are written in component form as 
\be
S^{\rm gen.}_e\ {\rm at}\ {\cal O}(e)  = (S^0_e + S^{\rm red.}_e)\ {\rm at}\ {\cal O}(e) 
\ee
with 
\ba
S^0_e\ {\rm at}\ {\cal O}(e) 
\A = \A \int d^2 x \ e \ \bigg\{ i v_{0a} \bar\chi \gamma^a \nu 
- \epsilon^{ij} v_0^a B^i \partial_a B^j 
+ \bar\lambda_0^i \chi B^i + \epsilon^{ij} \bar\lambda_0^i \nu B^j 
\nonu
\A \A 
- {1 \over 2} D_0 (B^i)^2 
+ {1 \over 2} A_0 (\bar\chi \chi + \bar\nu \nu) 
- \phi_0 \bar\chi \gamma_5 \nu \bigg\}, 
\label{S0-e}
\\
S^{\rm red.}_e\ {\rm at}\ {\cal O}(e) 
\A = \A \int d^2 x \ e \ \{ - 2 C (F^i)^2 + 2 (\bar\Lambda^i \chi F^i - \epsilon^{ij} \bar\Lambda^i \nu F^j) 
\nonu
\A \A 
+ A_0 B^i F^i - 2 M^{ij} B^i F^j + \cdots \}, 
\label{Sred-e}
\ea
where $S^0_e$ and $S^{\rm red.}_e$ mean the actions written in terms of 
only the fields for the minimal off-shell vector supermultiplet 
and in terms of the fields including the redundant (subsidiary) components, respectively. 
In Eqs.(\ref{S0-e}) and (\ref{Sred-e}) gauge invariant quantities \cite{ST4,WB} 
are denoted by 
\be
(A_0, \phi_0, F_{0ab}, \lambda_0^i, D_0) 
\equiv (M^{ii}, \phi, F_{ab}, \lambda^i + i \!\!\not\!\partial \Lambda^i, D + \Box C), 
\label{gauge-invariant}
\ee
with $F_{0ab} = \partial_a v_{0b} - \partial_b v_{0a}$, 
$F_{ab} = \partial_a v_b - \partial_b v_a$. 
The quantities (\ref{gauge-invariant}) are invariant ($v_{0a} = v_a$ transforms as an Abelian gauge field) 
under a SUSY generalized gauge transformation, $\delta_g {\cal V} = \Lambda^1 + \alpha \Lambda^2$, 
with an arbitrary real parameter $\alpha$ and generalized gauge parameters $\Lambda^i$ 
in the form of the $N = 2$ scalar superfields, 
and the fields $(A_0, \phi_0, v_{0a}, \lambda_0^i, D_0)$ constitute 
the minimal off-shell vector supermultiplet \cite{ST4}. 

By directly substituting the SUSY invariant relations (\ref{VSUSYinv}) and (\ref{SSUSYinv}) 
into the actions (\ref{S0-e}) and (\ref{Sred-e}), we obtain 
\ba
\A \A 
S^0_e(\psi)\ {\rm at}\ {\cal O}(e) \equiv \int d^2 x \left\{ 
{1 \over 4} e \kappa \xi (\xi^i)^2 \bar\psi^j \psi^j\bar\psi^k \psi^k \right\}, 
\label{S0-epsi}
\\
\A \A 
S^{\rm red.}_e(\psi)\ {\rm at}\ {\cal O}(e) \equiv \int d^2 x \left\{ 
- {1 \over 4} e \kappa \xi (\xi^i)^2 \bar\psi^j \psi^j\bar\psi^k \psi^k \right\}, 
\label{Sred-epsi}
\ea
which show Eq.(\ref{gauge-vanish}) by means of nontrivial cancellations 
among the four NG fermion self-interaction terms (the condensations of $\psi^i$) 
(\ref{S0-epsi}) and (\ref{Sred-epsi}). 

As for the relation between the $N = 2$ NLSUSY action (\ref{NLSUSYaction}) 
and the $N = 2$ LSUSY QED action for the minimal off-vector supermultiplet 
$(A_0, \phi_0, v_{0a}, \lambda_0^i, D_0)$, 
the terms (\ref{Sred-epsi}) for the redundant (subsidiary) structure of $S^{\rm gen.}_{N = 2{\rm SUSYQED}}$ 
can be absorbed into terms, ${1 \over 2} (F^i)^2$, in Eq.(\ref{gauge}) 
by defining the generalized (relaxed) SUSY invariant relations for the auxiliary fields $F^i$ \cite{ST3}, 
\be
F'^i(\psi) = F^i(\psi) - {1 \over 4} e \kappa^2 \xi \xi^i \bar\psi^j \psi^j \bar\psi^k \psi^k 
\label{gen-F}
\ee
with $F^i(\psi)$ in Eq.(\ref{SSUSYinv}). 
Then, the general SUSY QED action (\ref{SQEDaction}) reduces to 
the LSUSY QED action for the minimal off-shell vector supermultiplet in NL/L SUSY relation, i.e. 
\be
S^{\rm gen.}_{N = 2{\rm SUSYQED}}(\psi) 
= S^0_{N = 2{\rm SUSYQED}}(\psi) \vert_{F \rightarrow F'} 
+ [{\rm surface\ terms}] 
= S_{N = 2{\rm NLSUSY}}, 
\label{SQED0-NLSUSY}
\ee
when $\xi^2 - (\xi^i)^2 = 1$. 
The $S^0_{N = 2{\rm SUSYQED}}$ in Eq.(\ref{SQED0-NLSUSY}) means the local $U(1)$ gauge invariant 
and $N = 2$ LSUSY QED action given by \cite{ST3} 
\be
S^0_{N = 2{\rm SUSYQED}} 
= S^0_{{\cal V}{\rm kin}} + S^0_{{\cal V}{\rm FI}} 
+ S^0_{\Phi{\rm kin}} \vert_{F \rightarrow F'} + S^0_e 
\label{SQED0action}
\ee
with 
\ba
S^0_e = \A \A \int d^2 x \bigg[ \ 
e \ \bigg\{ i v_{0a} \bar\chi \gamma^a \nu 
- \epsilon^{ij} v_0^a B^i \partial_a B^j 
+ \bar\lambda_0^i \chi B^i 
+ \epsilon^{ij} \bar\lambda_0^i \nu B^j 
- {1 \over 2} D_0 (B^i)^2 
\nonu
\A \A 
+ {1 \over 2} A_0 (\bar\chi \chi + \bar\nu \nu) 
- \phi_0 \bar\chi \gamma_5 \nu \bigg\} 
+ {1 \over 2} e^2 (v_{0a}{}^2 - A_0^2 - \phi_0^2) (B^i)^2 \ \bigg]. 
\label{gaction}
\ea

Therefore, from Eqs.(\ref{SQED-NLSUSY}) and (\ref{SQED0-NLSUSY}) 
we can show the relations for the $N = 2$ SUSY QED action in the $N = 2$ NL/L SUSY relation, 
\be
S_{N = 2{\rm NLSUSY}} = S^{\rm gen.}_{N = 2{\rm SUSYQED}} 
= S^0_{N = 2{\rm SUSYQED}} \vert_{F \rightarrow F'} + [{\rm surface\ terms}], 
\label{NLSUSY-SQED}
\ee
without imposing a priori any special gauge conditions for the general gauge superfield (\ref{VSF}). 

We summarize our results as follows. 
In this letter we have argued systematically about the relation between the $N = 2$ NLSUSY model 
and the $N = 2$ SUSY QED theory in $d = 2$ superfield formulation. 
Based on the SUSY invariant relations (\ref{VSUSYinv}) and (\ref{SSUSYinv}) 
which are obtained from the (simplest) SUSY invariant constraints 
(\ref{constraintI}) and (\ref{constraintII}), 
we have attributed straightforwardly the general $N = 2$ SUSY QED action (\ref{SQEDaction}) 
to the $N = 2$ NLSUSY action (\ref{NLSUSYaction}) in Eq.(\ref{SQED-NLSUSY}). 
By studying the relation (\ref{gauge-vanish}) in the explicit component form, 
we have found that it vanishes due to the nontrivial cancellations 
among the four NG fermion self-interaction terms (the condensations of $\psi^i$) 
(\ref{S0-epsi}) and (\ref{Sred-epsi}). 
Since the generalization (relaxation) of the SUSY invariant relations for $F^i$ in Eq.(\ref{gen-F}) 
absorbes Eq.(\ref{Sred-epsi}) for the redundant (subsidiary) structure of the action (\ref{SQEDaction}) 
into the $N = 2$ LSUSY QED action (\ref{SQED0action}) for the minimal off-shell vector supermultiplet 
as in Eq.(\ref{SQED0-NLSUSY}), 
the relations for the $N = 2$ SUSY QED action in the $N = 2$ NL/L SUSY relation 
has been shown in Eq.(\ref{NLSUSY-SQED}) 
without imposing a priori any special gauge conditions for the general gauge superfield (\ref{VSF}). 

The similar results are anticipated in $d = 4$ and the investigation is important, 
and the extension of the arguments in NL/L SUSY relation to large $N$ is crucial in the SGM scenario. 
The SUSY Yang-Mills extension of the present arguments is an interesting problem.

\newpage

%
\newcommand{\NP}[1]{{\it Nucl.\ Phys.\ }{\bf #1}}
\newcommand{\PL}[1]{{\it Phys.\ Lett.\ }{\bf #1}}
\newcommand{\CMP}[1]{{\it Commun.\ Math.\ Phys.\ }{\bf #1}}
\newcommand{\MPL}[1]{{\it Mod.\ Phys.\ Lett.\ }{\bf #1}}
\newcommand{\IJMP}[1]{{\it Int.\ J. Mod.\ Phys.\ }{\bf #1}}
\newcommand{\PR}[1]{{\it Phys.\ Rev.\ }{\bf #1}}
\newcommand{\PRL}[1]{{\it Phys.\ Rev.\ Lett.\ }{\bf #1}}
\newcommand{\PTP}[1]{{\it Prog.\ Theor.\ Phys.\ }{\bf #1}}
\newcommand{\PTPS}[1]{{\it Prog.\ Theor.\ Phys.\ Suppl.\ }{\bf #1}}
\newcommand{\AP}[1]{{\it Ann.\ Phys.\ }{\bf #1}}

\end{document}